\newif\ifpdf
\ifx\pdfoutput\undefined
\pdffalse 
\else
\pdfoutput = 1 
\pdftrue
\fi

\documentclass [11pt, a4paper, openright, oneside] {article}

\usepackage {amsmath}
\usepackage {graphics}
\usepackage {graphicx}
\usepackage {longtable}
\usepackage {pgf,tikz}
\usepackage {cite} 
\usepackage {float,afterpage} 
\usepackage {verbatim} 
\usetikzlibrary{arrows,automata,shapes,snakes,backgrounds,topaths,petri}
\usepackage{subfigure}

\title{Parameter inference and model selection in signaling pathway models}
\date{}
\author{Tina Toni, Michael P. H. Stumpf}

\begin{document}

\maketitle


\abstract{To support and guide an extensive experimental research into systems biology of signaling pathways, increasingly more mechanistic models are being developed with hopes of gaining further insight into biological processes. In order to analyse these models, computational and statistical techniques are needed to estimate the unknown kinetic parameters. This chapter reviews methods from frequentist and Bayesian statistics for estimation of parameters and for choosing which model is best for modeling the underlying system. Approximate Bayesian Computation (ABC) techniques are introduced and employed to explore different hypothesis about the JAK-STAT signaling pathway.}

\section{Introduction}

It is crucial for cells to be able to sense the environment and react to changes in it. This is done through signaling pathways, which involve complex networks of often non-linear interactions between molecules. These interactions transduce a signal from outside the cell to trigger a functional change within a cell. Signaling pathways are important for differentiation, survival and adaptation to varying external conditions. The dynamics of these pathways are currently a subject of extensive experimental and computational research \cite{Klipp:2006p3726, Neves:2002p2243, Levchenko:2003p20089, Cho:2003p383, Papin:2005p10540}; and some of the most important biological signaling pathways have received considerable attention from mathematical modellers and theoretical systems biologists. These signaling networks include MAPK and Ras-Raf-ERK \cite{Fujioka:2006p20351, Apgar:2008p19125, Markevich:2004p20137, Babu:2004p19341, SureshBabu:2006p19868, Conzelmann:2004p19382, Kolch:2005p19657, Andrec:2005p19374, Schoeberl:2002p2523}, JAK-STAT \cite{Aaronson:2002p17948, Swameye:2003p130, BalsaCanto:2008p19037}, GPCR \cite{Modchang:2008p19045}, NF-$\kappa$B \cite{Cho:2003p383, Yue:2006p19056}.  It has become abundantly clear, however, that these signaling pathways can not be separated  from one another, but that they interact; this phenomenon is known as crosstalk \cite{Schwartz:1999p20164}. \\
\\
\\
A series of modeling approaches have been applied to the study of signaling pathways. Most widely used are ordinary differential equation (ODE) models that follow mass action kinetics. Also Boolean and Bayesian networks and Petri nets have been employed for modeling and simulation. The rich formalism underlying these different approaches has provided us with efficient tools for the analysis of signaling models. \\
\\
Computational approaches have been mainly used for simulation and for studying qualitative properties of signaling pathways such as motifs and feedback loops \cite{Tyson:2003p10542, Bhalla:1999p2519}, and quantitative properties such as signal duration, signal amplitude and amplification \cite{Heinrich:2002p10523, SaezRodriguez:2004p19354, JulioVera:2008p17580}.\\
\\
To develop and utilize detailed quantitative signaling models we require the values of all the parameters, such as kinetic rates of protein turnover and post-translational modifications (e.g. phosphorylation or dimerization). Due to technological restrictions and cost  it is impossible to measure all the parameters experimentally. In this chapter we review computational tools that can be used for parameter inference for ODE models. While many studies have dealt with the subject of parameter estimation, relatively little attention has been given to model selection; that is, which model(s) to use for inference. Despite this, "What is the best model to use?" is probably the most critical question for making valid inference from the data \cite{Burnham:2002p4089}, and this is the second topic that we touch on in this chapter.\\
\\ 
There are two broad schools of thought in statistical inference: frequentist and Bayesian. In frequentist statistics one talks about point estimates and confidence intervals around them. The likelihood function is a central concept in statistical inference, and is used in both frequentist and Bayesian settings. It equals the probability of the data given the parameters, and it is a function of parameters,  
$$
L(\theta) = P(D|\theta).
$$
The cannonical way of obtaining the point estimate is by taking a maximum likelihood estimate; i.e. the set of parameters for which the probability of observing the data is highest. On the other hand, Bayesian statistics is based on probability distributions. Here one aims to obtain the posterior probability distribution over the model parameters, which is proportional to the product of a suitable prior distribution (which summarizes the user's prior knowledge or expectations) and the likelihood (the information that is obtained from the data). \\
\\
In the following sections we will review how frequentist and Bayesian statistics can be used to estimate parameters of ODE models of signaling pathways, and how to choose which model has the highest support from the data. We then outline an approximate Bayesian computation algorithm based on Sequential Monte Carlo and apply it to the JAK-STAT signaling pathway where we will illustrate aspects related to parameter estimation and model selection. 

\section{Parameter inference}

Signaling pathway models include numerous parameters, and it is generally impossible to obtain all of these values by experimental measurements alone. Therefore parameter inference (also referred to as model calibration, model fitting or parameter estimation by different authors) algorithms can be used to  estimate these parameter values computationally. A variety of different approaches have been developed and are being used used; they all share the two main ingredients: a \textit{cost function}, which reflects and penalizes the distance between the model and experimental data, and an \textit{optimization algorithm}, which searches for parameters that optimize the cost function. The most commonly used cost functions in a frequentist approach are the likelihood (one wants to maximize it) and the least squares error (one wants to minimize it). The Bayesian equivalent  to a cost function is the Bayesian posterior distribution.\\
\\
There are many different kinds of optimization algorithms. Their goal is to explore the landscape defined by cost function and find the optimum (i.e. minimum or maximum, depending on the type of cost function used). The simplest are the local gradient descent methods (e.g. Newton's method, Levenberg-Marquardt). These methods are computationally fast, but are only able to find local optima. When the cost function landscape is complex, which is often the case for signaling models with high dimensional parameter space, these methods are unlikely to find the global optimum, and in this case more sophisticated methods need to be used. Multiple shooting \cite{Peifer:2007p3987} performs better in terms of avoiding getting stuck in local optima, but, as argued by Brewer \textit{et al.} \cite{Brewer:2007p3554} may perform poorly when measurements are sparse and data are noisy. A large class of optimization methods are global optimization methods that try to explore complex surfaces as widely as possible; among these genetic algorithms are particularly well known and have been applied to ODE models\cite{JulioVera:2008p17580}. Moles \textit{et al.} \cite{Moles:2003p8} tested several global optimization algorithms on a $36$-parameter biochemical pathway model and showed that the best performing algorithm was a stochastic ranking evolutionary strategy \cite{Runarsson:2000p689} (software available \cite{Ji:2006p390, Zi:2006p131}). Further improvements in computational efficiency of this algorithm were obtained by hybrid algorithms incorporating local gradient search and multiple shooting methods \cite{RodriguezFernandez:2006p381, BalsaCanto:2008p19037}.\\
\\
To obtain an ensemble of good parameter values, an approach based on simulated annealing \cite{Kirkpatrick:1983p611} and Monte Carlo search though parameter space can be used \cite{Brown:2003p727, Brown:2004p726}. In a Bayesian setting, MCMC methods \cite{Vyshemirsky:2008p14865} (software available \cite{Vyshemirsky:2008p19927}) and unscented Kalman filtering  \cite{Quach:2007p19250} have been applied to estimate the posterior distribution of parameters. Bayesian methods do not only estimate confidence intervals, but provide even more information by estimating of the whole posterior parameter distribution. To obtain confidence intervals for a point estimate in a frequentist setting, a range of techniques can be applied that include variance-covariance matrix based techniques \cite{Bard:1974p11303}, profile likelihood \cite{Venzon:1988p11240} and bootstrap methods \cite{Efron:1993p11400}.\\ 
\\
Parameter estimation should be accompanied by identifiability and sensitivity analyses. If a parameter is non-identifiable, this means it is difficult or impossible to estimate due to either model structure (structural non-identifiability) or insufficient amount or quality or data measurements (statistical non-identifiability) \cite{Hengl:2007p20238, Schmidt:2008p20239, Yue:2006p19056}. Structurally non-identifiable parameters should ideally be removed from the model. Sensitivity analysis studies how model output behaves when varying parameters \cite{Saltelli:2008p14297}. If model output changes a lot when parameters are varied slightly, we say that the model is sensitive to changes in certain parameter combinations. Recently, the related concept of sloppiness has been introduced by Sethna and co-workers \cite{Brown:2003p727, Gutenkunst:2007p728}. They call a model sloppy when the parameter sensitivity eigenvalues are roughly evenly distributed over many decades; those parameter combinations with large eigenvalues are called sloppy and those with low eigenvalues stiff. Sloppy parameters are hard to infer and carry very little discriminatory information about the model. The concepts of identifiability, sloppiness and parameter sensitivity are, of course, related:  non-identifiable parameters and sloppy parameters are hard to estimate precisely because they can be varied a lot without having a large effect on model outputs; the corresponding parameter estimates will thus have large variances. A parameter with large variance can, in a sensitivity context, be interpreted as one to which the model is not sensitive if the parameter changes.  

\section{Model selection}

Model selection methods strive to rank the candidate models, which represent hypothesis about the underlying system, relative to each other according to how well they explain the experimental data. Crucially, the chosen model is not the "true" model, but the best model from the set of candidate models. It is the one which we should probably use  for making inferences from the data. Generally, the more parameters are included in the model, the better a fit to the data can be achieved. If the number of parameters equals the number of data points, there is always a way of setting the parameters so that the fit will be perfect. This is called overfitting. Wel \cite{Wel:1975p12810} famously addressed a question of "how many parameters it takes to fit an elephant", which practically suggests that if one takes a sufficiently large number of parameters, a good fit can always be achieved. The other extreme is underfitting, which results from using too few parameters or too inflexible a model. A good model selection algorithm should follow the principle of parsimony, also referred to as Occam's razor, which aims for to determine the model with the smallest possible number of parameters that adequately represents the data and what is known about the system under consideration.\\
\\
The probably best known method for model selection is (frequentist) hypothesis testing. If ODE models are nested (i.e. one model can be obtained from the other by setting some parameter values to zero), then model selection is generally performed using the likelihood ratio test \cite{Swameye:2003p130, Timmer:2004p474}. If both models have the same number of parameters and if there is no underlying biological reason to choose one model over the other, then we choose the one which has a higher maximum likelihood. However, if the parameter numbers differ, then the likelihood ratio test penalizes overparameterization. \\
\\
If the models are not nested, then model selection becomes more difficult but a variety of approaches have been developed that can be applied in such (increasingly more common) situations.  Bootstrap methods \cite{Efron:1993p11400, Timmer:2004p474} are based on drawing many so-called bootstrap samples from the original data by sampling with replacement, and calculating the statistic of interest (e.g. an achieved significance level of a hypothesis test) for all of these samples. This distribution is thin compared to the real data. \\
\\
Other model selection methods applicable to non-nested models are based on information-theoretic criteria \cite{Burnham:2002p4089} such as the Akaike Information Criteria (AIC) \cite{Akaike:1973p12727, Akaike:1974p3502, Swameye:2003p130, Timmer:2004p474}. These methods involve a goodness-of-fit term and a term measuring  the parameteric complexity of the model. The purpose of this complexity term is to penalize models with high number of parameters; the criteria by which this term should be chosen can differ considerably among the various proposed measures. \\
\\
In a Bayesian setting model selection is done through so-called Bayes factors (for comprehensive review see \cite{Kass:1995p2898}). We consider two models, $m_1$ and $m_2$ and would like to determine which model explains the data $x$ better. The Bayes factor measuring the support of model $m_1$ compared to model $m_2$, is given by
$$
B_{12} = \frac{p(x|m_1)}{p(x|m_2)} = \frac{\int p(x|m_1,\theta_1) p(\theta_1|m_1) d \theta_1}{\int p(x|m_2, \theta_2)p(\theta_2|m_2) d \theta_2}.
$$
To compute it, marginal likelihoods have to be computed, and this is done by integrating non-linear functions over all possible parameter combinations. This can be a challenging problem when the dimension of the parameter space is high, and Vyshemirsky \textit{et al.} asses various methods how this can be done efficiently \cite{Vyshemirsky:2008p14865}. A Bayesian version of information-theoretic model selection techniques introduced above is the Bayesian Information Criterion (BIC) \cite{Schwarz:1978p6945, Brown:2003p727}, which is an approximation of the logarithm of the Bayes factor. Unlike the AIC, which tends towards overly complex models as the data saturates, the BIC chooses correct models in the limit of infinite data availability.\\
\\
There are several advantages of Bayesian model selection compared to traditional hypothesis testing. Firstly, the models being
compared do not need to be nested. Secondly, Bayes factors do not only weigh the evidence against a hypothesis (in our
case $m_2$), but can equally well provide evidence in favour of it. This is not the case for traditional hypothesis testing where a small $p$-value only indicates that the
null model has insufficient explanatory power. However, one cannot conclude from a large $p$-value that the two models are equivalent, or that the
null model is superior, but only that there is not enough evidence to distinguish between them. In other words, ``failing to reject'' the null hypothesis cannot be translated to ``accepting'' the null hypothesis \cite{Cox:1974p15668,Kass:1995p2898}. Thirdly, unlike the posterior probability of the model, the $p$-value does not provide any direct interpretation of the weight of
evidence (the $p$-value is not the probability that the null hypothesis is true). We expect that Bayesian methods will also deal better with so-called sloppy parameters because they are based on explicit marginalization over model parameters. 

\section{Approximate Bayesian computation techniques}

When formulating the likelihood for an ODE model, one normally assumes the Gaussian error distribution on the data points: by definition this is the only way of defining a likelihood for a deterministic model. Moreover, it might be hard to analytically work with the likelihood (e.g. finding maximum likelihood estimate and integrating the marginal probabilities). Approximate Bayesian computation (ABC) methods have been conceived with the aim of inferring posterior distributions by circumventing the use of the likelihood, in favour of exploiting the computational efficiency of modern simulation techniques by replacing calculation of the likelihood with a comparison between the observed data and simulated data. These approaches are also straightforwardly applied to ODE model of signaling networks.\\
\\
Let $\theta$ be a parameter vector to be estimated. Given the prior distribution $p(\theta)$, the goal is to approximate the posterior
distribution, $p(\theta|x) \propto f(x|\theta)p(\theta)$, where $f(x|\theta)$ is the likelihood of $\theta$ given the data $x$. ABC methods have the following generic form:
\begin{description}

 	\item [1] Sample a candidate parameter vector $\theta^*$ from some proposal distribution $p(\theta)$.

	\item [2] Simulate a data set $x^*$ from the model described by a conditional probability distribution $f(x|\theta^*)$.

	\item [3] Compare the simulated data set, $x^*$, to the experimental data, $x_0$, using a distance function, $d$, and tolerance $\epsilon$; if $d(x_0, x^*)
	\leq \epsilon$, accept $\theta^*$. The tolerance $\epsilon \geq 0$ is the desired level of agreement between  $x_0$ and $x^*$.
\end{description}
The output of an ABC algorithm is a sample of parameters from a distribution $p(\theta|d(x_0, x^*) \leq \epsilon)$. If $\epsilon$ is sufficiently small then the distribution $p(\theta|d(x_0, x^*) \leq \epsilon)$ will be a good approximation for the ``true'' posterior distribution, $p(\theta|x_0)$. \\
\\
The most basic ABC algorithm outlined above is known as the ABC rejection algorithm; however, recently more sophisticated and computationally efficient ABC methods have been developed. They are based on Markov chain Monte Carlo (ABC MCMC) \cite{Marjoram:2003p5} and Sequential Monte Carlo (ABC SMC) techniques \cite{Sisson:2007p2, Toni:2009p20998, Beaumont:2008p21781, Moral:2009p20996}. They have recently been applied to dynamical systems modeled by ordinary differential equations and stochastic master equations; ABC SMC has been developed  for approximating the posterior distribution of the model parameters and for model selection using Bayes factors \cite{Toni:2009p20998}. In the next section we illustrate the use of ABC SMC for parameter estimation and model selection in the context of the JAK-STAT signaling pathway. 

\section{Application to JAK-STAT signaling pathway}

\begin{figure}[h]	
	\begin{center}
	\includegraphics[width = 125mm]{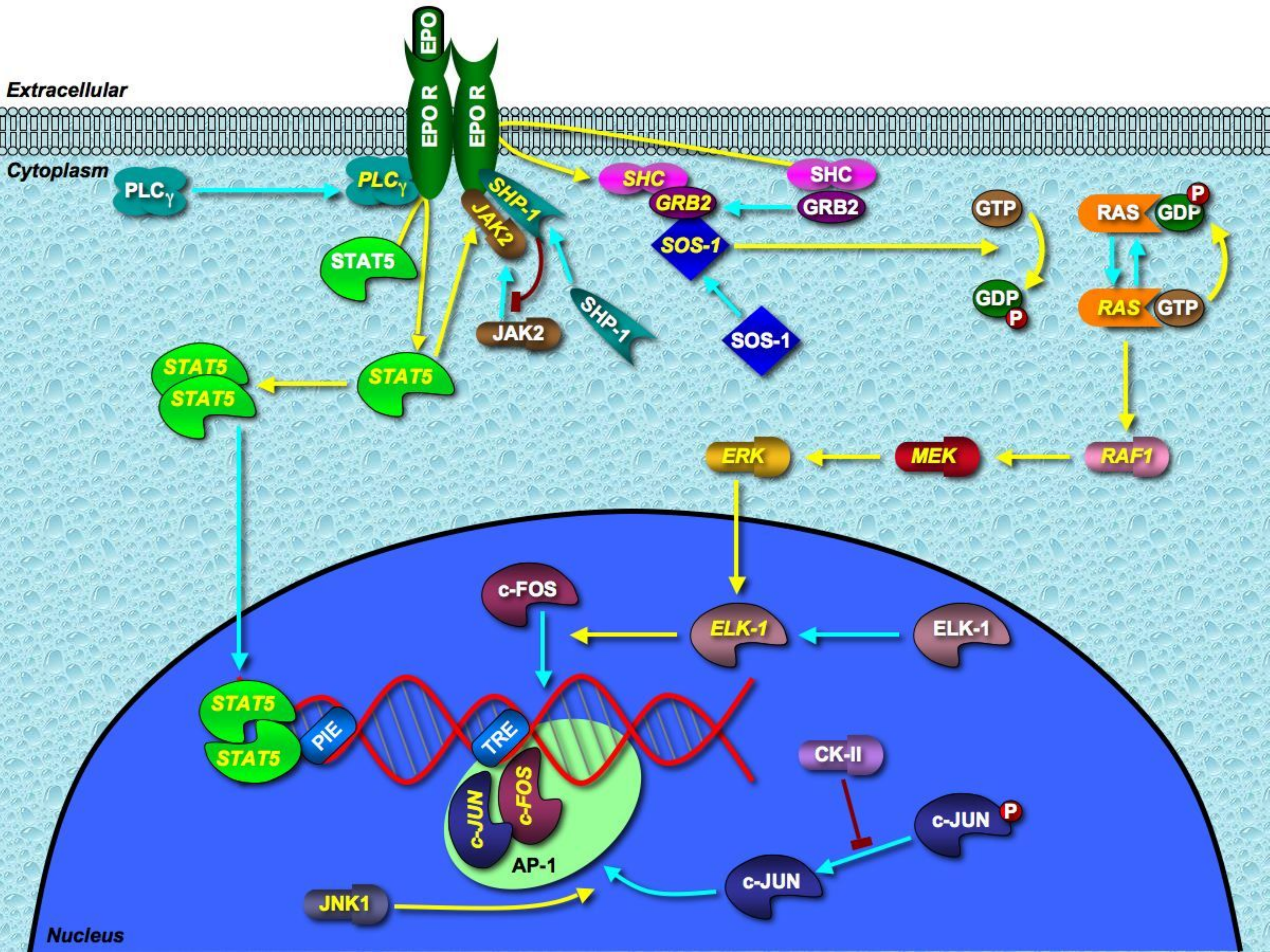}
	\end{center}
	\caption{\label{fig:epo} \small{STAT5 signalling pathway.}}
\end{figure}

\begin{figure}[h]	
	\begin{center}
	\includegraphics[width = 120mm]{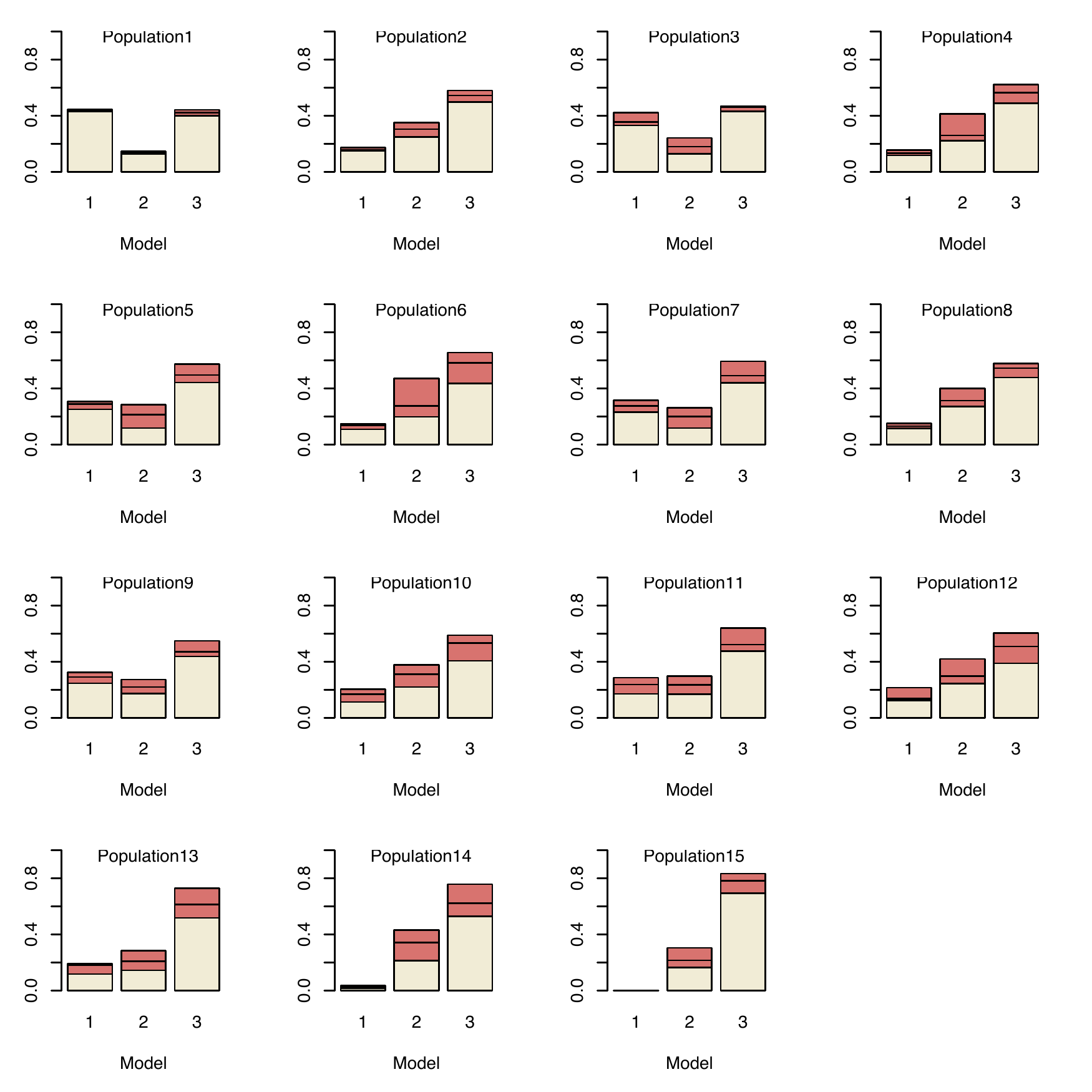}
	\end{center}
	\caption{\label{fig:model_selection} \small{Histograms show populations of the model parameter $m$. Population 13 represents the approximation of the marginal posterior distribution of $m$. The red sections present 25\% and 75\% quantiles around the median.}}
\end{figure}

The JAK-STAT signalling pathway is involved in signalling through several surface receptors and STAT proteins, which act as signal transducers and activators of transcription \cite{Darnell:1997p2005, Horvath:2000p1959}.  Here we look at models of signalling though the erythropoietin receptor (EpoR), transduced by STAT5 (Figure \ref{fig:epo}). Signalling through this receptor is crucial for proliferation, differentiation, and survival of erythroid progenitor cells \cite{Klingmuller:1996p1955}.\\
\\
When the Epo hormone binds to the EpoR receptor, the receptor's cytoplasmic domain becomes phosporylated, which creates a docking site for signalling molecules, in particular the transcription factor STAT5. Upon binding to the activated receptor, STAT5 first becomes phosphorylated, then dimerizes and translocates to the nucleus, where it acts as a transcription factor. There have been competing hypotheses about what happens with STAT proteins at the end of the signalling pathway. Originally it had been suggested that STAT proteins get degraded in the nucleus in an ubiquitin-asssociated way \cite{Kim:1996p1953}, while other evidence suggests that they are dephosphorylated in the nucleus and then transported back to the cytoplasm \cite{Koster:1999p1954}. \\
\\
Here we want to understand how STAT5 protein transduces the signal from the receptor in the membrane through the cytoplasm into the nucleus. We have approached this problem by applying ABC SMC for model selection and parameter estimation to data collected for the JAK-STAT signaling pathway. The most suitable model from model of a STAT5 part of the JAK-STAT signaling pathway among the three proposed models was chosen and parameters have been estimated.\\
\\
The ambiguity about the shutoff mechanism of  STAT5 in the nucleus triggered the development of several mathematical models \cite{Swameye:2003p130, Muller:2004p265, Timmer:2004p474}, each describing a different hypothesis. These models were then fitted to experimental data and systematically compared to each other using statistical methods of model selection. The model selection procedure ruled in favour of a cycling model, where STAT5 reenters the cytoplasm.\\
\\
Timmer \textit {et al.} \cite{Swameye:2003p130, Muller:2004p265, Timmer:2004p474} developed a continuous mathematical model for STAT5 signalling pathyway, comprising of four differential equations. They assume mass action kinetics and denote the amount of activated Epo-receptors by $EpoR_A$, monomeric unphosphorylated STAT5 molecules by $x_1$, monomeric phosphorylated STAT5 molecules by $x_2$, dimeric phosphorylated STAT5 in the cytoplasm by $x_3$ and dimeric phosphorylated STAT5 in the nucleus by $x_4$. 
The most basic model Timmer \textit{et al.} developed, under the assumption that phosphorylated STAT5 does not leave the nucleus, consists of the following kinetic equations:
\begin{eqnarray}
	\dot{x}_1 &=& - k_1 x_1 EpoR_A \label{eq:basic_1} \label{first} \\
	\dot{x}_2 &=& - k_2 x_2^2 + k_1 x_1 EpoR_A \nonumber \\
	\dot{x}_3 &=& - k_3 x_3 + \frac{1}{2}k_2x_2^2 \nonumber \\
	\dot{x}_4 &=& k_3 x_3  \label{eq:basic_4}. \label{fourth}
\end{eqnarray}
One can then assume that phosphorylated STAT5 de-dimerizes and leaves the nucleus and this can be modelled by adding appropriate kinetic terms to the equations (\ref{first}) and (\ref{fourth}) of the basic model:
\begin{eqnarray*}
	\dot{x}_1 &=& - k_1 x_1 EpoR_A + 2 k_4 x_4\\
	\dot{x}_4 &=& k_3 x_3 - k_4 x_4.
\end{eqnarray*}
Timmer \textit{et al.} develop their cycling model further by assuming a delay in moving of STAT5 out of the nucleus. They write ODE equations for $x_1$ and $x_4$ for this model as
\begin{eqnarray}
	\dot{x}_1 &=& - k_1 x_1 EpoR_A + 2 k_4 x_3(t-\tau)\\
	\dot{x}_4 &=& k_3 x_3 - k_4 x_3(t-\tau),
\end{eqnarray}
while equations for $x_2$ and $x_3$ remain as above. The outcome of their statistical analysis is that this model fits the data best, which leads them to the conclusion that this is the most appropriate model.\\
\\
Instead of Timmer's chosen model, we propose a similar model with clear physical interpretation. Instead with $x_3(t-\tau)$, we propose to model the delay of phosphorylated STAT5 $x_4$ in the nucleus with $x_4(t-\tau)$: 
\begin{eqnarray*}
	\dot{x}_1 &=& - k_1 x_1 EpoR_A + 2 k_4 x_4(t-\tau)\\
	\dot{x}_4 &=& k_3 x_3 - k_4 x_4(t-\tau).
\end{eqnarray*}
We have performed the ABC SMC model selection algorithm \cite{Toni:2009p20998} on the following models: (1) Cycling delay model with $x_3(t-\tau)$, (2) Cycling delay model with $x_4(t-\tau)$, (3) Cycling model without a delay. The model parameter $m$ can therefore take values 1, 2 and 3. \\
\\
Figure \ref{fig:model_selection} shows intermediate populations leading to the approximation of the marginal posterior distribution of $m$ (population 15). Bayes factors can be calculated from the last population and according to the conventional interpretation of Bayes factors  \cite{Kass:1995p2898}, it can be concluded that there is very strong evidence in favour of models 2 and 3 compared to model 1. However, there is only moderate evidence for model 3 being more suitable than model 2.
\\

\section{Discussion}
Modeling biological signaling or regulatory systems requires reliable parameter estimates.But the experimental dissection of signaling pathways is costly and laborious;  it furthermore seems unreasonable to believe that the same set of parameters describes a system across all possible environmental, physiological and developmental conditions. We are therefore reliant on efficient and reliable statistical and computational methods in order to estimate parameters and, more generally, reverse engineer mechanistic models.\\
\\
As we have argued above, any such estimate must include a meaningful measure of uncertainty. A rational approach to modeling such systems should furthermore allow for the comparison of competing models in light of available data. The relative new ABC approaches are  able to meet both objectives. Furthermore as we have shown elsewhere they are not limited to deterministic modelling approaches but are also readily applied to explicitly stochastic dynamics; in fact it is possible to compare the explanatory power of deterministic and stochastic dynamics in the same mechanistic model.   \\
\\
One of the principal reasons for applying sound inferential procedures in the context of dynamical systems is to get a realistic appraisal of the robustness of these systems. If, as has been claimed, only a small set of parameters determines the system outputs then we have to ascertain these with certainty. It is here, in the reverse engineering of potentially sloppy dynamical systems, where the Bayesian perspective may be most beneficial. 

\footnotesize



\end{document}